\documentclass[useAMS,usenatbib]{mn2e}

\usepackage{float} % If comment this, figure moves to Page 2
\usepackage{epsfig}
\usepackage{graphicx}
\usepackage[latin1]{inputenc}
\usepackage{latexsym}
\def\simlt{\ \raise -2.truept\hbox{\rlap{\hbox{$\sim$}}\raise5.truept   %
\hbox{$<$}\ }}
\def\simgt{\ \raise -2.truept\hbox{\rlap{\hbox{$\sim$}}\raise5.truept   %
\hbox{$>$}\ }}                                                          %

\def\be{\begin{equation}}
\def\ee{\end{equation}}
\def\newline{\hfil\break}

\def\la{\mathrel{\hbox{\rlap{\hbox{\lower4pt\hbox{$\sim$}}}\hbox{$<$}}}}
\def\ga{\mathrel{\hbox{\rlap{\hbox{\lower4pt\hbox{$\sim$}}}\hbox{$>$}}}}

\def\mach{\mathcal{M}}
\title[The $P_{1.4}-\mach$ correlation for radio relics]{The correlation between radio power and Mach number for radio relics in galaxy clusters}
\author[Colafrancesco, Marchegiani and Paulo]{S. Colafrancesco$^{1}$\thanks{E-mail:sergio.colafrancesco@wits.ac.za}, P. Marchegiani$^{1}$ and C.M. Paulo$^{1,2}$\\
$^{1}$School of Physics, University of the Witwatersrand, Private Bag 3, 2050-Johannesburg, South Africa\\
$^{2}$Departamento de F\'{\i}sica, Faculdade de Ci{\^e}ncias, Universidade Eduardo Mondlane, Po Box 257, Maputo, Mo\c{c}ambique\\
}

\begin{document}

\date{Accepted: 2017 July 13. Received: 2017 July 09; in original form: 2017 May 09.}

\pagerange{\pageref{firstpage}--\pageref{lastpage}} \pubyear{2014}

\maketitle

\label{firstpage}

\begin{abstract}
We discuss a new technique to constrain models for the origin of radio relics in galaxy clusters using the correlation between the shock Mach number and the radio power of relics. This analysis is carried out using a sample of relics with information on both the Mach numbers derived  from X-ray observation,  $\mach_X$, and using spectral information from radio observations of the peak and the average values of the spectral index along the relic, $\mach_R$.
We find that  there is a lack of correlation between $\mach_X$ and $\mach_R$;  
this result is an indication that the spectral index of the relic is likely not due to the acceleration of particles operated by the shock but it is related to the properties of a fossil electrons population.
We also find that the available data on the correlation between the radio power $P_{1.4}$ and Mach numbers ($\mach_R$ and $\mach_X$) in relics indicate that neither the DSA nor the adiabatic compression can simply reproduce the observed $P_{1.4}-\mach$ correlations.
Furthermore, we find that the radio power is not correlated with $\mach_X$, whereas it is not possible to exclude a correlation with $\mach_R$.
This also indicates that the relic power is mainly determined by the properties of a fossil electron population rather than by the properties of the shock.
Our results require either to consider models of shock (re)acceleration that go beyond the proposed scenarios of DSA and adiabatic compression at shocks, or to reconsider the origin of radio relics in terms of other physical scenarios. 
\end{abstract}


%%%%%%%%%%%%%%%%%%%%%%%%%%%%%%%%%%%%%%%%%%%%%%%
\section{Introduction}     

Galaxy clusters  are  the largest gravitationally-bound structures in the universe. They are also the largest storage rooms for dark matter, baryonic material in the form of stars and galaxies, and diffuse  gas  in the form of a  hot ($T \, \sim \, 10^{7} - 10^{8}$ K) and tenuous ($n \, \sim \, 10^{-3} - 10^{-2}$ cm$^{-3}$) ionized plasma (intra cluster medium, ICM) that fills the space between galaxies.
Beyond the baryonic material of the ICM and galaxies, many clusters host non-thermal components in the form of $\sim \mu$G magnetic fields and relativistic particles (i.e., cosmic rays electrons) as indicated by the steep-spectrum, diffuse radio emission due to synchrotron emission of these relativistic particles.

During the last decade considerable progress has been made in the observations and in the physical description of diffuse radio sources in galaxy clusters which are referred to as radio halos (RH), radio mini-halos (RMH) and radio relics (RR), depending on their location, morphology, and polarization properties (see, e.g., Feretti et al. 2012 for a review).
 
Radio halos found in the most massive clusters have spatial morphologies similar to their ICM X-ray emission (e.g. Deiss et al. 1997; Govoni et al. 2001a), even if in some cases a difference is noted (e.g. Govoni et al. 2012), and are usually centrally located in galaxy clusters that are experiencing merging processes (e.g. Giovannini, Tordi \& Feretti 1999), even though in some cases no evidence of major mergers but only accretion interactions with galaxy groups have been found (see, e.g., Ogrean \& Br\"uggen 2013 for the case of the Coma cluster). 
RHs show little (or no) synchrotron polarization ($\leq$ 5\%, Feretti et al. 2012), although a few RHs with regions showing more sensitive polarized radio emission have been reported (e.g. Govoni et al. 2005).
 
Radio mini-halos are also typically located at cluster center but with their radio emission concentrated on smaller scales ($\sim$ 100-500 kpc) (see, e.g., Ferrari et al. 2008 for a review) of the central radio galaxy; they have steep spectral index and are found often in relaxed, cool-core clusters, with or without signs of dynamical  activity (see, e.g., Cassano, Gitti \& Brunetti 2008 and references therein).

Radio relics are typically observed towards the cluster periphery at $\sim$ Mpc distances from the cluster centre (e.g. Feretti et al. 2012). They are observed to be roundish or elongated (with the major axis roughly perpendicular to the direction of the cluster center, see, e.g., Bonafede et al. 2009a; van Weeren et al. 2009; Giovannini et al. 2013) and seem to be strongly polarized, with fractional linear polarization observed at 1.4 GHz lying above 10\%, and reaching values up to 50\% in some specific regions (e.g. Bonafede et al. 2009a).\\
Based on their morphology these radio sources are usually divided into two main groups: "radio gischt", with a very elongated morphology, that include also double-relics with the two relics located on different sides of a cluster centre, and "radio phoenixes" with more roundish shapes.
RRs have usually steep synchrotron spectra ($\alpha_r \sim $ 1.0--1.4, where the flux density is $S_\nu \, \propto \, \nu^{-\alpha_r}$), they are characterized by low-surface brightness emission ($\sim \, \mu$Jy  arcsec$^{-2}$ at 1.4 GHz), they have large physical extents (up to $\sim \, $ 1-1.5 Mpc), they are highly-polarized, and they have 1.4 GHz radio powers in the range of $10^{23}-10^{26}$ W Hz$^{-1}$ (e.g. Feretti et al. 2012).

Radio relics have been initially considered as the remnants of radio galaxy lobes (Giovannini, Feretti \& Stanghellini 1991). 
Subsequently, the claim that relics seem to trace the shock fronts produced by accretion activity or merging events (see, e.g., En\ss lin et al. 1998) led to consider intra-cluster shocks as the origin of radio relics. 
The idea of accelerating particles via diffusive shock acceleration (DSA) was widely considered (see, e.g., En\ss lin et al. 1998; Kang \& Ryu 2011) thus attributing the relics origin primarily to a particle acceleration mechanism, even though also the different mechanism of adiabatic compression of clouds of fossil relativistic electrons and magnetic fields is in place in these structures (see, e.g., En\ss lin \& Gopal-Krishna 2001).\\
More detailed studies (Kempner et al. 2004) led to consider that the "radio gischt" are produced by shock waves originated in the ICM, and that the "radio phoenix" are remnants of AGN activity, therefore suggesting that only "radio gischt" have to be properly considered as radio relics. 
However,  there have been recently a growing number of cases where also the "radio gischt" have been shown to be connected with radio galaxies (see, e.g., the cases of  the cluster PLCKG287.0+32.9, Bonafede et al. 2014; the Bullet cluster, Shimwell et al. 2014, 2015; the cluster  A3411-3412, van Weeren et al. 2017a), suggesting therefore that this could be a common origin for all radio relics. 
Nonetheless, in the case of the cluster PLCKG287.0+32.9, given the large size of the relic -- larger than the distance that the electrons can cross in a magnetized medium without loosing their energy -- and the complex structure of the relic, Bonafede et al. (2014) suggested that more than one radio galaxy could produce the electron distribution associated to the relic.

As for the physical mechanism producing radio relics, models based on shock acceleration of particles from the thermal pool via DSA have been strongly challenged by both observational and theoretical arguments. In fact, not all clusters showing the presence of merging host a radio relic, and DSA has been found to be very inefficient for weak shocks, as it is usually found in galaxy clusters (see Kang, Ryu \& Jones 2012). In addition, in a number of cases where it was possible to measure the Mach number of the shock from X-ray observations, it has been found that the Mach number value was not sufficient to produce, via DSA, a radio spectral index of the order of the observed one (see, e.g., the cases of  Coma, Ogrean \& Br\"uggen 2013; CIZA J2242.8+5301, Ogrean et al. 2013a; see also Akamatsu et al. 2017 for the most recent compilation). In some other cases, the position of the shock is also not coincident with the position of the radio relic (see, e.g., the case of the Bullet cluster, Shimwell et al. 2015).

These problems led to require the presence of a fossil population of non-thermal electrons, possibly produced by pre-existing lobes/relics of radio galaxies, that can be then reaccelerated at the shock front (Pinzke, Oh \& Pfrommer 2013; Bonafede et al. 2014). This assumption could reduce the efficiency of the shock acceleration required to produce the relic, and could then explain why not all merging clusters host a relic.\\
%.
However, even the inclusion of a pre-existing electron population does not solve all the problems connected with the DSA interpretation of the relic origin, like the lack of detection of galaxy clusters in the gamma-rays which strongly challenges the role of DSA as the origin of the electron acceleration, even though the electrons are part of a pre-existing fossil population (Vazza \& Br\"uggen 2014), and the fact that DSA models do not predict a sharp decrease of the radio spectrum at high frequencies as observed in some relic (Kang \& Ryu 2015).

Another scenario that has been proposed is the electron re-acceleration by turbulences produced behind the shock fronts, that can explain the observed value of the spectral index of the relic in the Toothbrush cluster, that is flatter than the value predicted by the DSA, as well as the shift in the positions of the relic and the shock front in the same cluster (Fujita et al. 2015). 
However, also this model has the problem that turbulences should also accelerate protons (Vazza \& Br\"uggen 2014; Vazza et al. 2016), thus generating a quite strong gamma-ray emission.

In this framework, alternative explanations of relics being produced by, e.g., adiabatic compression of lobes of radio galaxies deserves to be further considered as the origin for radio relics.
Van Weeren et al. (2017a) recently disfavoured such a possibility because of the observed flattening of the spectral index close to the position of the shock in the merging cluster A3411-A3412. However, En\ss lin \& Gopal-Krishna (2001) previously showed that a change of the spectral index depending on the time elapsed between the injection of the electrons and the compression operated by the shock is indeed expected in this model, because not only the electrons density but also the magnetic field is compressed (and amplified) by the shock, and in some cases this effect can produce a flattening of the radio spectral index.
The same authors also pointed out that a compression model can be favoured w.r.t. an acceleration model because shock waves can have difficulties in accelerating the electrons inside a bubble of relativistic plasma because the sound speed inside such a bubble is much higher than in the surrounding ICM and therefore a shock wave with a given velocity is seen with a much lower Mach number by the electrons inside the bubble.
Moreover, Pfrommer \& Jones (2011) studied a model where a radio galaxy with one or more lobes crosses a shock front, finding that the action of the compression of the lobes can produce a morphology of the resulting radio emission quite similar to those observed in relics.

The observational and theoretical arguments presented so far indicate that the physical origin of radio relics in clusters is not clear yet: a connection between radio relics and radio galaxies is suggested in some clusters, but deserves to be thoroughly and further investigated in all clusters hosting radio relics in order to understand if it can be considered as a general mechanism for the formation of relics. 
In addition, it is not clear yet if all the sources presently classified as relics have indeed the same origin; in some case, in fact, it has been found that some sources classified as relics are not really relics, but are simply residual of radio galaxies or AGN activity, as in the cases of A13 (Juett et al. 2008; George et al. 2017).
Other open problems include the reasons why relics are usually found at the periphery of clusters, what is the origin of the bridge between the relic and the radio halo observed in some cluster (e.g., Coma, see Brown \& Rudnick 2011), and if the polarized filaments observed within the radio halos in some clusters (e.g., A2255, Govoni et al. 2005; MACS J0717.5+3745, Bonafede et al. 2009b) are actually relics seen projected close to the direction of the cluster center (see, e.g., Pizzo et al. 2011).\\
Once all the relics are analyzed systematically and are properly classified according to their morphology and their connection with radio galaxies, it is then necessary to proceed with a detailed analysis of their properties in order to understand which scenario for their formation is favoured.

In this paper, we focus on the  the connection between Mach numbers (representing shock acceleration properties) and relic radio power measured at $1.4$ GHz. We derive these data  for a sample of  20 relics in 16 galaxy clusters where we have both radio and X-ray information about the relics and the possible associated intra-cluster shocks. 
We describe in Sect. 2 the theoretical framework of the relic radio emission in the model of particle acceleration  via DSA and in the adiabatic compression of fossil relativistic electrons, focussing in particular on the relation between radio power and shock Mach number we can expect in the two scenarios. We derive the observational correlation between the relic radio luminosity at 1.4 GHz $P_{1.4}$ and their Mach number $\mach$ in Sect. 3 analyzing the correlation for the Mach numbers derived from radio spectral information and those derived from X-ray observations. We discuss our results in Sect. 4, and we  summarize our conclusions in the Sect. 5 of this paper.\\
A flat, vacuum dominated cosmology with $\Lambda CDM$ cosmology with $H_{0} = 67.3$ km s$^{-1}$ Mpc$^{-1}$, $\Omega_\Lambda = 0.68$ and $\Omega_M = 0.32$ is assumed throughout this paper.

%%%%%%%%%%%%%%%%%%%%%%%%%%%%%%%%%%

\section{Theoretical framework}

We describe in this section the basic information about the theory of shock acceleration and of magnetic field compression at relics that are necessary to derive the theoretical predictions for the $P_{1.4}-\mach$ correlation we want to study.

\subsection{Acceleration of particles via DSA}

The DSA theory (Drury 1983; Blandford \& Eichler 1987) is based on the original idea of Fermi (1949), according to which a particle crossing a shock surface moving at a velocity $v$ is accelerated by a factor $\Delta E/E \sim v/c$; since in a galaxy cluster the shock waves move in a magnetized plasma, a particle diffusing in the plasma irregularities can cross several times the shock surfaces, increasing in this way its energy. Bell (1978) showed that the spectrum of particles produced in this way is depending only on the shock strength, and it is expected to take the shape of a power law, $N(p)\propto p^{-s_i}$, where $p$ is the momentum of the particle, and
\begin{equation}
s_i=\frac{2u_2+u_1}{u_1-u_2} \;,
\end{equation}
where $u_1$ and $u_2$ are the upstream and downstream velocities in the shock rest frame, respectively. 
Therefore, for strong shock with compression ratio $r=u_1/u_2>>1$, a spectrum with $s_i \sim 1$ is expected, while for a weak shock with $r \simgt 1$ the spectral index becomes very steep ($s_i\rightarrow \infty $). 
In a gas with a polytropic index $\gamma_g = 5/3$, and in the strong shock limit, the compression ratio takes the value $r=(\gamma_g+1)/(\gamma_g-1)=4$, implying that the flattest possible value of the spectral index is $s_i = 2$. In term of the shock Mach number, defined as $\mach = v / c_s$, where $c_s$ is the sound speed in the medium, the compression ratio for a polytropic index $\gamma_g = 5/3$ is given by
\begin{equation}
r=\frac{4\mach^2/3}{\mach^2/3+1} \;,
\end{equation}
and this translates in the relation
\begin{equation}
s_i=2\frac{\mach^2+1}{\mach^2-1} \;.
\end{equation} 
A pressure jump of the thermal gas is also expected:
\begin{equation}
\frac{P_2}{P_1}=\frac{4r-1}{4-r} \;,
\end{equation}
where $P_1$ and $P_2$ are the upstream and downstream pressures, respectively (see, e.g., Landau \& Lifshitz 1959). 

Since the radio spectral index is connected with the equilibrium electrons spectral index by $\alpha_r = (s-1)/2$ (see, e.g., Longair 1994), where $F_\nu \propto \nu^{-\alpha_r}$, the expected radio spectral index can be connected to the shock strength; the Mach number of the shock required to produce a radio spectrum with spectral index $\alpha_r$ is thus given by
\begin{equation}
\mathcal{M}^2 = {2 \alpha_r +3 \over 2\alpha_r-1}
\label{mach.radio.1}
\end{equation}
if the electrons are located close to the shock and do not have time to reach the equilibrium with the energy losses ($s=s_i$), and is given by
\begin{equation}
\mathcal{M}^2 = {\alpha_r +1 \over \alpha_r-1}
\label{mach.radio.2}
\end{equation}
if the electrons have reached the equilibrium with the energy losses ($s=s_i+1$). 
As a consequence, DSA predicts radio spectral index that must be $\simgt 0.5$ close to the shock front and $\simgt 1$ far away from the shock front in the strong shock limit.

Numerical studies have been performed to estimate the strength of the shocks produced during the evolution of a galaxy cluster, considering both the shocks produced in the continuous growth of matter falling in the cluster gravitational potential well (accretion shocks) and the shocks produced during the merging of different pre-existing clusters (merging shocks). External shocks, produced in regions outside the virialized radius, where the matter usually has not yet been heated by other shocks, are expected to have very high Mach numbers, even up to $\mach \sim100$ (e.g. Ryu et al. 2003), because the sound speed in these regions is lower due to the low temperature of the plasma, whereas internal shocks are expected to have usually lower Mach numbers ($\mach<4$, Ryu et al. 2003), but are expected to be more effective in accelerating particles, because they are located in regions with higher densities of seed electrons. The efficiency of acceleration of particles by shocks with low Mach number is still not well understood, but in general it is supposed to be low (see, e.g., Hoeft \& Br\"uggen 2007).

If the seed electrons population is not the thermal bath of the cluster ICM, but a previous population of relativistic particles that are losing their energy, for weak shocks DSA is not expected to change heavily the spectrum of the electrons at low energies, because the spectrum of the electron is expected to maintain the shape provided by the seed population at low energies, and to assume the shape predicted by the DSA at higher energies (see, e.g., Fig.8 in Macario et al. 2011). However, by increasing the strength of the shock the spectrum of the electrons will approach the shape produced via DSA along all the spectrum of the electrons (Kang \& Ryu 2011). Therefore, we should expect a sudden change of the electrons properties for shocks with $\mach \sim 4$ or higher.

A complementary model involves the turbulences that can be produced during the shock and can reaccelerate stochastically the electrons (see Fujita et al. 2015). 
This model is often considered as a complementary mechanism of direct acceleration models that is able to maintain the electrons accelerated in a wider region and for a longer time after the shock passage differently from the standard DSA (e.g. Kang, Ryu \& Jones 2017).

\subsection{Adiabatic compression of fossil relativistic electrons}

The theory of adiabatic compression of clouds of fossil relativistic electrons operated by shock waves has been developed by En\ss lin \& Gopal-Krishna (2001). The time dependent equation describing the evolution of the electrons spectrum, when there is not an active source of particles and the the diffusion is neglected, is given by:
\begin{equation}
-\frac{dp}{dt}=\frac{4}{3}\frac{\sigma_T}{m_e c} u_R p^2 + \frac{1}{3} \frac{1}{V} \frac{dV}{dt} p \;,
\end{equation}
where $p$ is the normalized momentum of the particle, $u_R$ is the energy density of the radiative fields (CMB photons and magnetic field) that produce energy losses, and $V$ is the volume of the cloud of relativistic plasma. By defining the compression ratio as
\begin{equation}
C(t)=\frac{V_0}{V(t)} \;,
\end{equation}
it is possible to calculate the momentum of a particle with initial momentum $p_0$ at a time $t$:
\begin{equation}
p(p_0,t)=\frac{p_0}{C(t)^{-1/3}+p/p_*(t)} \;,
\label{eq.momentum_change}
\end{equation}
where the characteristic momentum is given by
\begin{equation}
\frac{1}{p_*(t)}=\frac{4}{3}\frac{\sigma_T}{m_e c} \int_{t_0}^t dt' u_R(t')\left(\frac{C(t')}{C(t)}\right)^{1/3} \;.
\end{equation}
For an initial power-law shape
\begin{equation}
N(p)=N_0 p^{-s}
\label{eq.powerlaw}
\end{equation}
between $p_{min}$ and $p_{max}$, the spectrum at time $t$ is given by
\begin{equation}
N(p,t)=N_0 C(t)^{(s+2)/3} p^{-s} (1-p/p_*(t))^{s-2}
\end{equation}
between $p(p_{min},t)$ and $p(p_{max},t)$. Due to the effect of compression, also the magnetic field energy density is changed by a factor
\begin{equation}
u_B(t)=u_{B,0} (V/V_0)^{-4/3}
\label{eq.bcompress}
\end{equation}
and, since $u_B \propto B^2$, this means that $B(t) \propto C(t)^{2/3}$.

In the model of En\ss lin \& Gopal-Krishna (2001), the clouds of plasma are injected by radio galaxies, expand by losing energy due to radiative losses and adiabatic expansion until pressure equilibrium with the environment is reached, then remain in a state where electrons only lose energy by radiative losses at a low rate until the cloud is eventually compressed by a shock wave. Given the higher sound speed inside the cloud compared to the environment, the shock is expected to have a very low Mach number inside the cloud, and therefore it is not expected to accelerate the electrons via DSA, but just by compressing the cloud. The compression factor can be calculated from the pressure jump:
\begin{equation}
C=\left(\frac{P_2}{P_1}\right)^{3/4} \;,
\label{eq.compr.factor}
\end{equation}
where $P_1$ and $P_2$ are the upstream and downstream pressures, respectively. In this stage the cloud is expected to be compressed along the direction of the shock wave motion, and to expand perpendicularly to this direction, assuming the typical elongated shape of radio relics. After this stage, the electrons are subject to energy losses and their radio emission drops quickly.

We note that the eq.(\ref{eq.powerlaw}) is in general an approximation of the real electron spectrum that is expected in a cloud of fossil electrons; in fact, because of the energy losses of the electrons, the spectrum is expected to change with the distance from the source, and this effect can be parametrized with an exponential cutoff at high energies, with a cutoff energy value that is changing across the cloud producing a steepening of the spectrum (see, e.g., van Weeren et al. 2017a for the case of A3411-A3412). However, the effect of the compression is to re-energize the electrons and to amplify the magnetic field (see eqs.\ref{eq.momentum_change} and \ref{eq.bcompress} respectively), and as a consequence the electrons that are producing the radio emission at a given frequency are located in a lower energy region of the spectrum w.r.t. the electrons in the initial cloud (see discussion in the next subsection), and in this lower energy region the effect of the exponential cutoff is expected to be less strong. Therefore, whereas the full structure of the spectrum of the electrons in the cloud is necessary to understand the detailed structure of a specific relic, for a statistical study that involves the average properties of a number of relics the approximation of a single power law spectrum as in eq.(\ref{eq.powerlaw}) should not impact heavily on the final results.

\subsection{Radio emission}

Electrons accelerated by the previous mechanisms are expected to produce radio emission by synchrotron emission when interacting with the magnetic field. Hoeft \& Br\"uggen (2007) calculated the radio power that is produced by DSA:
\begin{equation}\label{eq:epr}
P_{\nu} \propto S \cdot n_{d} \cdot \xi_{e} \cdot \nu^{-\alpha_r} \cdot T_{d}^{3/2}\, \frac{B^{1+\alpha_r}}{(B_{CMB}^2+B^{2})} \Psi(\mathcal{M})  \;,
\end{equation}
where $S$ is the shock surface area, $n_{d}$ is the downstream electron density, $\xi_{e}$ is the electron acceleration efficiency, $T_{d}$ is the downstream electron temperature, $B$ is the relic magnetic field, and $\Psi(\mathcal{M})$ is the function that includes the dependence of the radio power from the Mach number. 
According to Fig.4 of Hoeft \& Br\"uggen (2007), the dependence of $P_{\nu}$ on the Mach number has a very steep increase in the weak shock region ($\mathcal{M}<4$), where the slope of the function $\Psi(\mach)$ is found to be close to 10--11 for $\mach\simlt3$, while it is almost flat in the strong shock region ($\mathcal{M}>4$). Therefore, we can expect a very quick drop in the relic radio emission for $\mach\simlt4$.

\begin{figure}
\centering
\begin{tabular}{c}
\includegraphics[width=85mm,height=65mm]{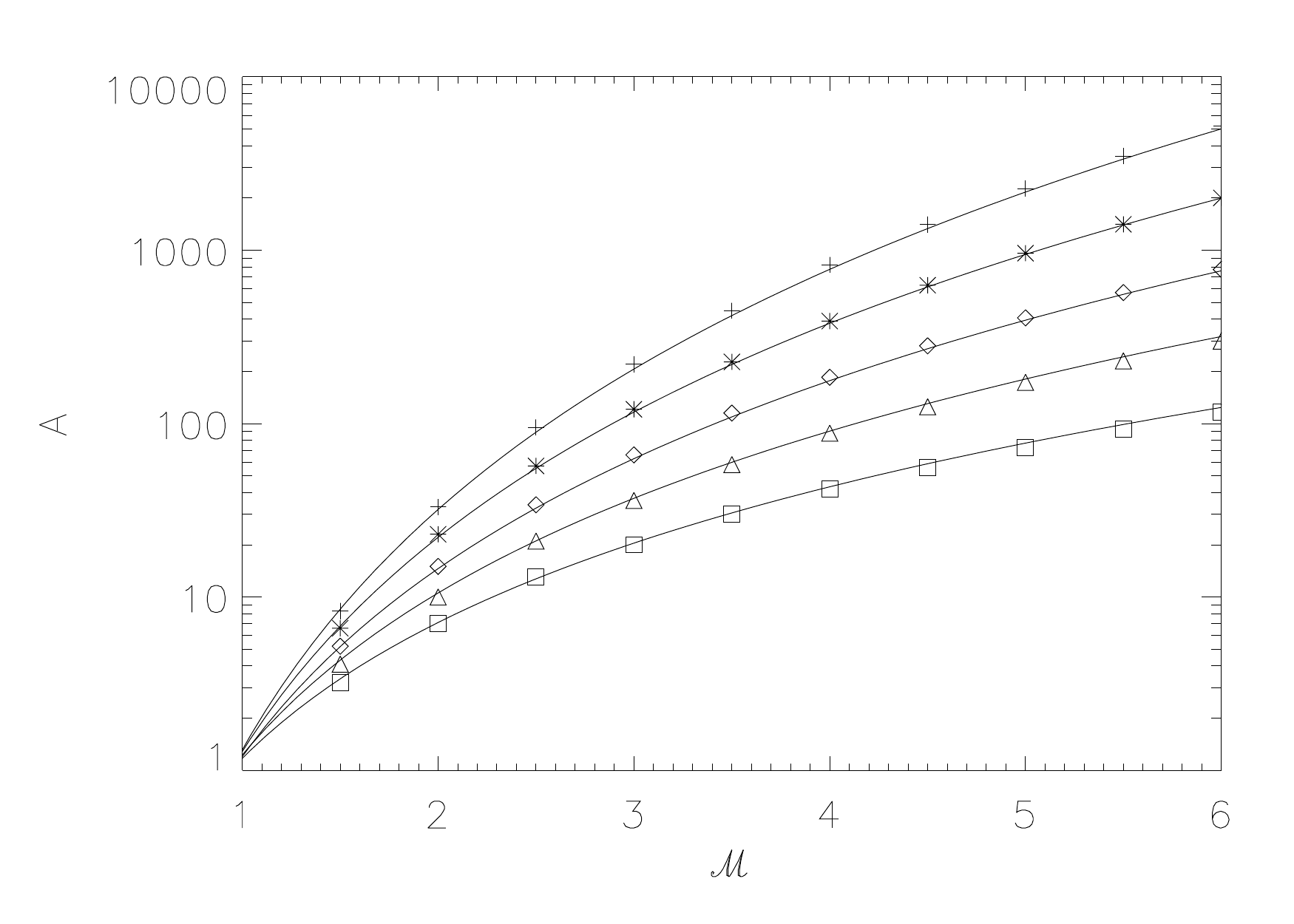}
\end{tabular}
\caption{Amplification factor $A$ calculated according to eq.(\ref{amplif.compr}) for values of the eletrons spectral index $s=2.5,3.0,3.5,4.0,4.5$ from the bottom to the top as a function of the Mach number; the solid lines are proportional to $\mach^{s+0.1}$.}
\label{fig:amp}
\end{figure}

On the other side, the adiabatic compression model allows to have a boosting of the radio emission during the compression phase even for low Mach number values (En\ss lin \& Gopal-Krishna 2001). 
By writing the radio luminosity as:
\begin{equation}
P_\nu \propto N \frac{B^{1+\alpha_r}}{(B_{CMB}^2+B^{2})} V
\label{eq.power.compr}
\end{equation}
we can find that, for a magnetic field smaller than $B_{CMB}\sim3$ $\mu$G that does not impact on energy losses, the radio emission is amplified by a factor
\begin{equation}
A\propto C^{(s+2)/3} \times [C^{2/3}]^{1+\alpha_r} \times C^{-1} \;.
\label{amplif.compr}
\end{equation}
For example, for a shock with $\mach\sim3$, we can estimate a compression ratio of $r=3$, and a pressure ratio of $P_2/P_1=11$. This corresponds to a compression factor $C\sim6$. Therefore, the magnetic field is amplified by a factor $\sim3.3$ according to eq.(\ref{eq.compr.factor}), and the electron density, for a spectral index $s=3.5$ (that in the adiabatic compression scenario is not determined by the shock strength but by the radio galaxy jet), is amplified by a factor $\sim27$, while the volume is reduced by a factor $C$. Therefore, for the corresponding radio spectral index $\alpha_r=1.25$, the radio power is amplified by a factor $\sim27\times3.3^{2.25}\times 6^{-1}\sim 66$. 
In Figure \ref{fig:amp}, we show this amplification factor calculated for several values of the Mach number and of the electrons spectral index $s$; we find that the adiabatic compression model does not predict a sudden change in the radio luminosity around $\mach\sim4$, but a continuous increase of the amplification factor following a power law with $A\propto\mach^{s+0.1}$. 

Another effect is a possible flattening of the radio spectrum, because if the electrons spectrum is ageing after the injection episode, we can expect that the spectrum is steepening towards high energies (e.g. Sarazin 1999); at a given frequency, the energy of the electrons emitting at that frequency is $(E/\mbox{GeV})^2\sim (\nu/16\mbox{ MHz}) (B/\mu\mbox{G})^{-1}$ (Longair 1994). Therefore, when  the magnetic field is increasing by compression, the energy of the electrons that are emitting at a given frequency is smaller, and the electrons spectrum at lower energy can be flatter. In the case we considered previously, the energy of the electrons emitting at a given frequency is smaller by a factor $3.3^{-1/2}\sim0.55$. 
According to eq.(\ref{eq.momentum_change}) we can see that another effect of the adiabatic compression is to change the energy of the electrons by a factor of the order of $C^{1/3}$, therefore producing a shift toward high energies of the electrons spectrum and a flattening of the radio spectrum. 
We note that the electrons that in the cloud have already lost almost all their energy (in the high-energy part of the spectrum) are not reaccelerated in this scenario, therefore we should expect a sudden and strong decrease of the radio emission at high frequency.

The compression mechanism also produces an alignment of the magnetic field lines: this increases the polarization level of the radio emission, and produces an anisotropy of the radio emission (that is proportional to the component of the magnetic field perpendicular to the line of sight). If the direction of the shock is almost perpendicular to the line of sight, we should then observe a relic with a very elongated shape in the direction perpendicular to the shock, with high luminosity and high polarization; if the direction of the shock is close to the observer direction, we will observe a more roundish shape (the compression is happening along the line of sight), with a lower luminosity and a lower degree of polarization. Therefore, in this scenario both the "radio gischt" and the "radio ghosts" can have the same origin, but different geometrical orientations.

Therefore, the adiabatic compression scenario should produce, unlike the DSA one: 
\textit{i)} a distribution of the radio luminosity as a function of the Mach number that does not show a sudden change around $\mach=4$; 
\textit{ii)} a high frequency steepening of the radio spectrum much stronger than in the DSA scenario; 
\textit{iii)} a correlation between the shape, the luminosity, and polarization level of the relic (relics with a very elongated shape should appear more luminous and more polarized); 
\textit{iv)} since electrons lose their energy on a time-scale $\simlt1$ Gyr after the injection, the shock front must be located very close to the relic, otherwise the adiabatic compression is no longer able to produce a detectable radio emission, because it is not reaccelerating the electrons.

These considerations suggest that it is necessary to perform a study of the correlation between the luminosity of the relic and the observed Mach number of the shock, as well as an examination of the spectral and spatial properties of the relics and their morphological connection to radio galaxies to discriminate between the acceleration and the adiabatic compression scenario. In this paper we will focus on the first topic, and will address the second one in a future paper (Colafrancesco et al., in preparation).

\section{Correlation between radio luminosity and Mach number}

\subsection{The cluster sample}

As a preliminary work to the present paper, we build a sample of 60 galaxy clusters with detected radio relics extracted from the literature. 
In this sample, a total of 77 radio relics were found; among them 43 are considered as single relics (i.e., one relic within the cluster) and 17 are considered double relics (i.e., two relics located on different sides of a cluster centre). 
We will discuss in details the statistical properties of this sample and the morphological analysis of each one of these cases in a forthcoming paper (Colafrancesco et al., in preparation), while we will focus here on the sample of clusters for which we have information on Mach numbers derived both through radio and X-ray observations in order to study their correlation with the relic radio power.

The determination of the Mach number of a shock associated to a radio relic is not a trivial operation. Several different methods have been used in the literature, and sometimes they give quite different results for the same shocks. The different proposed methods are based on X-ray data (through the density or temperature jump; e.g. Markevitch \& Vikhlinin 2007), on the radio spectral index assuming DSA (e.g. Blandford \& Eichler 1987), on the Sunyaev-Zeldovich effect (through the pressure jump; e.g. Erler et al. 2015), on the ellipticity of the relic (through the ratio between the major and minor axis as produced by shock compression; e.g. Pfrommer \& Jones 2011), and on the level of polarization (e.g. Kierdorf et al. 2017). At present, determinations of the Mach number from X-ray and from the radio spectral index values are the most used methods, and we will refer to them in the following analysis.

From X-ray observations, the shock jump conditions are given by the Rankine-Hugoniot conditions (e.g. Landau \& Lifshitz 1959), according to which the Mach number, for a polytropic index $5/3$, is related to the density and to the temperature jump by:
\begin{eqnarray}
\frac{n_2}{n_1} & = & \frac{4\mach_X^2}{\mach_X^2+3} \label{eq.machx.1} \\
\frac{T_2}{T_1} & = & \frac{5\mach_X^4+14\mach_X^2-3}{16\mach_X^2} 
\label{eq.machx.2}
\end{eqnarray}
where we indicate with $\mach_X$ the Mach number that can be determined with this method. The determination of $\mach_X$ from the temperature jump is in principle more solid than other methods, because it is less subject to effects of projections along the line of sight (see e.g. discussion in Ogrean et al. 2013b), and also allows to understand if the jump is produced by a shock or by a cold front depending on the direction of the jump; however, it requires a higher number of photon counts to perform the fitting to the X-ray data, that usually is difficult to obtain in the external regions of galaxy clusters. 
The determination of $\mach_X$ from the jump in the X-ray surface brightness is a more solid technique but it shows the mentioned problems in interpreting the data, such as projection effects. The determination of $\mach_X$ via density jump is also known to be more accurate for low Mach numbers because of the different shapes of the two functions in eqs. (\ref{eq.machx.1}) and (\ref{eq.machx.2}) (e.g. Markevitch \& Vikhlinin 2007). 
In the following we will collect from the literature the results obtained with both these methods, and we will select the data that the authors consider more reliable, where the error bars are smaller.

The other method to determine the Mach number of a shock, which is usually used in literature, is related to the use of radio spectral index information. In particular, eq.(\ref{mach.radio.1}) should be used when the radio spectral index is measured close to the shock front, and the electrons spectrum is the injection one because they do not have time to lose their energy, while eq.(\ref{mach.radio.2}) should be used if the electrons have reached the equilibrium with the energy losses. Unfortunately, it is not always easy to determine if the electrons have reached the equilibrium, and as a consequence it is not clear to establish which equation must be used. Often, the first equation is used when the spectral index is smaller than 1, and the second equation when the spectral index is larger than 1. As a result, when working with a sample of clusters where the radio spectral index are not measured in a uniform way and the Mach numbers are not derived accordingly, it is possible to obtain paradoxical results where shocks with higher Mach numbers produce relics with steeper spectral index (see, e.g., table 4 in Hindson et al. 2014). 
Therefore, in the following we determined the Mach number from radio observations $\mach_R$ in two ways:
\begin{itemize}
\item from the flattest observed value of the radio spectral index, $\alpha_{peak}$, using eq.(\ref{mach.radio.1}) to determine $\mach_{R,p}$;
\item from the averaged value of the radio spectral index along the relic $\alpha_{av}$, using eq.(\ref{mach.radio.2}) to determine $\mach_{R,a}$.
\end{itemize}
While the first method is expected to be more precise in determining the shock properties because the electrons are expected to be captured close to the shock, it is more difficult to achieve because it is necessary to measure the spectral index just at the position of the shock, therefore requiring to identify with precision the position of the shock, and performing measurements with high spatial resolution that must be the same at all the frequencies where the radio emission is observed. 
The second method is experimentally more solid, but is less precise theoretically, because the average value of the spectral index along the relic will involve regions where the electrons have just been emitted, regions where the electrons are approaching the equilibrium, and possibly also regions where the effect of the acceleration is finished, and the electrons are ageing thus quickly steepening their spectrum; therefore, it is just an approximation to assume that the average spectral index is equal to the equilibrium one. 

We searched for the relics where the Mach number has been determined with both the X-ray and the radio spectral index methods. The results are shown in Table \ref{tab:Mach_X_radio}. 
In building this Table, we avoided to use values of the radio spectral index that can not be accounted by the DSA (i.e. $\alpha_{peak} \le 0.5$ and $\alpha_{av} \le 1$); when these kinds of data were presented in literature, we searched for other determination of the spectral index, and we increased the error bars to account for both values.

\begin{table*}
\centering
\caption{The cluster relic sub-sample: radio power $P_{1.4}$, X-ray derived Mach number values $\mach_X$ (from eqs.\ref{eq.machx.1}--\ref{eq.machx.2}), flattest radio spectral index $\alpha_{peak}$ and corresponding derived radio Mach number $\mach_{R,p}$ (from eq.\ref{mach.radio.1}), average spectral index $\alpha_{av}$ and derived Mach number $\mach_{R,a}$ (from eq.\ref{mach.radio.2}). References: (1) Shimwell et al. (2015); (2) Macario et al. (2011); (3) Kassim et al. (2001); (4) Giacintucci et al. (2008); (5) Bourdin et al. (2013); (6) Trasatti (2014); (7) Ogrean \& Br\"uggen (2013); (8) Giovannini et al. (1991); (9) Trasatti et al. (2015); (10) Clarke \& En\ss lin (2006); (11) van Weeren et al. (2012); (12) Johnston-Hollitt (2003); (13) Akamatsu et al. (2012a); (14) R\"ottgering et al. (1997); (15) Akamatsu \& Kawahara (2013); (16) Stroe et al. (2013); (17) Akamatsu et al. (2015); (18) Kale et al. (2012); (19) Akamatsu et al. (2012b); (20) van Weeren et al. (2011); (21) Owers et al. (2014); (22) Shimwell et al. (2016); (23) van Weeren et al. (2013); (24) van Weeren et al. (2017a); (25) Botteon et al. (2016a); (26) Govoni et al. (2001b); (27) Lindner et al. (2014); (28) Botteon et al. (2016b); (29) Orr\`u et al. (2007); (30) Eckert et al. (2016); (31) Pizzo \& De Bruyn (2009); (32) Akamatsu et al. (2017); (33) Bonafede et al. (2009b); (34) Ma, Ebeling \& Barrett (2009); (35) van Weeren et al. (2017b); (36) Ogrean et al. (2013b). Notes: \textit{a}: error bar not present in the cited paper, we assumed 0.1 on the last significant digit; \textit{b}: error bar assumed to match the value of $0.92\pm0.02$ given by Trasatti et al. (2015); \textit{c}: van Weeren et al. (2012) give the value $1.0\pm0.2$, but from the spectrum they publish it is clear that this spectral index does not match well the points at 1.7 and 2.3 GHz and is affected mainly from the point at 147 MHz that has a big error bar, therefore we fixed the value at the maximum value allowed by their error bar; \textit{d}: error bar assumed to match the value of $0.9\pm0.1$ estimated by Johnston-Hollitt (2003); \textit{e}: value and error estimated from the map published in Lindner et al. (2014); \textit{f}: value estimated from the full resolution data of Pizzo \& De Bruyn (2009) and error fixed to match the value they derive after correcting to have the same angular resolution at the different frequencies ($0.8\pm0.1$) and the value $1.4$ estimated by Feretti et al. (1997); \textit{g}: tentative estimate derived by us from the jump in the temperature observed between the regions A6 and A15 by Ma et al. (2009), that have temperatures of $19\pm8$ and $12\pm5$ keV respectively; \textit{h}: error bar not present in the cited paper, estimated by us from the errors on the temperature; \textit{i}: error bar not present in the cited paper, estimated to be of the same order than a previous case discussed in the same paper, before removing one point from the fitting; \textit{j}: in the cited paper the Mach number is given as $\mach_X\simgt3.5-4$.}
\label{tab:Mach_X_radio}
\begin{tabular}{c c c c c c c}
\hline
Cluster &  $P_{1.4}$ (W Hz$^{-1}$)  & $\mathcal{M}_X$  & $\alpha_{peak}$  & $\mach_{R,p}$ & $\alpha_{av}$  & $\mach_{R,a}$  \\
\hline
Bullet & $(2.3\pm0.1)\times10^{25}$  (1) & $2.5^{+1.3}_{-0.8}$  (1) & $0.8\pm0.1^a$  (1) & $2.8\pm0.4$ & $1.07\pm0.03$  (1) & $5\pm1$  \\
A754 & $(4.5\pm0.2)\times10^{22}$  (2) & $1.57^{+0.16}_{-0.12}$  (2) & $1.8\pm0.1^a$  (3) & $1.59\pm0.04$ & $2.02\pm0.04$  (2) & $1.72\pm0.02$ \\
A521 & $(2.8\pm0.2)\times10^{24}$ (4) & $2.42\pm0.19$ (5) & $1.0\pm0.1$ (4) & $2.2\pm0.2$ & $1.48\pm0.01$ (4) & $2.27\pm0.02$  \\
Coma & $(2.3\pm0.2)\times10^{23}$ (6) & $1.9^{+0.16}_{-0.40}$ (7) & $1.0\pm0.2$ (8) & $2.2\pm0.4$ & $1.21\pm0.03$ (8) & $3.2\pm0.2$ \\
A2256 & $(4.1\pm0.3)\times10^{24}$ (9) &  $1.7\pm0.2^h$ (9) & $0.6\pm0.1^a$ (9) & $5\pm2$ & $1.2\pm0.3^b$ (10) & $3\pm2$  \\
Toothbrush (N) & $(6.0\pm0.4)\times10^{25}$ (11) & $1.9^{+0.75}_{-0.42}$ (36) & $0.65\pm0.05$ (11) & $3.8\pm0.6$ & $1.10\pm0.02$ (11) & $4.6\pm0.4$  \\
Toothbrush (E) & $(1.8\pm0.2)\times10^{24}$ (11) & $2.4^{+0.46}_{-0.36}$ (36) & $1.0\pm0.1^a$ (11) & $2.2\pm0.2$ & $1.2\pm0.2^c$ (11) & $3\pm2$ \\
A3667 (NW) & $(2.9\pm0.2)\times10^{25}$ (12) & $2.41\pm0.39$ (13) & $0.7\pm0.1$ (12) & $3.3\pm0.8$ & $1.1\pm0.2^d$ (14) & $5\pm4$ \\
A3667 (SE) & $(2.4\pm0.2)\times10^{24}$ (12) & $1.75\pm0.13$ (15) & $0.9\pm0.1$ (12) & $2.4\pm0.3$ & $1.2\pm0.2$ (12) & $3\pm2$ \\
Sausage (N) & $(1.5\pm0.2)\times10^{25}$ (16) & $3.15\pm0.52$ (15) & $0.60\pm0.05$ (16) & $5\pm1$ & $1.06\pm0.04$ (16) & $6\pm2$  \\
Sausage (S) & $(2.2\pm0.2)\times10^{24}$ (16) & $1.7^{+0.4}_{-0.3}$ (17) & $0.8\pm0.1^a$ (16) & $2.8\pm0.4$ & $1.29\pm0.05$ (16) & $2.8\pm0.2$ \\
A3376 & $(6.0\pm0.5)\times10^{23}$ (18) & $2.94\pm0.60$ (19) & $1.0\pm0.2$ (18) & $2.2\pm0.4$ & $1.70\pm0.06$ (18) & $1.96\pm0.06$ \\
A2034 & $(2.8\pm0.8)\times10^{23}$ (20) & $1.59^{+0.06}_{-0.07}$ (21) & $0.9\pm0.1$ (22) & $2.4\pm0.3$ & $1.7\pm0.1$ (22) & $2.0\pm0.1$  \\
A3411 & $(3.0\pm0.2)\times10^{24}$ (23) & $1.7\pm0.2^i$ (24) & $0.9\pm0.1$ (24) & $2.4\pm0.3$ & $1.2\pm0.1$ (24) & $3.3\pm0.8$ \\
A115 & $(3.9\pm0.2)\times10^{24}$ (25) &  $1.7\pm0.1$ (25) & $-$ & $-$ & $1.1\pm0.5$ (26) & $5\pm11$ \\
El Gordo (NW) & $(3.1\pm0.9)\times10^{25}$ (27) & $4.1^{+3.4}_{-0.9}$ (28) & $0.86\pm0.15$ (27) & $2.56\pm0.45$ & $1.19\pm0.09$ (27) & $3.4\pm0.7$  \\
El Gordo (SE) & $(4.5\pm1.5)\times10^{24}$ (27) & $3.75\pm0.25^j$ (28) & $1.1\pm0.1^e$ (27) & $2.1\pm0.1$ & $1.4\pm0.1$ (27) & $2.4\pm0.3$  \\
A2744 & $(6.5\pm0.5)\times10^{24}$ (29) & $1.7^{+0.5}_{-0.3}$ (30) & $0.9\pm0.1$ (29) & $2.4\pm0.3$ & $1.1\pm0.1$ (29) & $5\pm2$ \\
A2255 & $(4.8\pm0.2)\times10^{23}$ (31) & $1.42^{+0.19}_{-0.15}$ (32) & $0.9\pm0.1$ (31) & $2.4\pm0.3$ & $1.1\pm0.3^f$ (31) & $5\pm7$ \\
MACS J0717.5+3745 & $(1.05\pm0.06)\times10^{26}$ (33) & $1.6\pm1.5^g$ (34) & $1.0\pm0.5$ (35) & $2.2\pm0.2$ & $1.3\pm0.1$ (33) & $2.8\pm0.4$ \\
\hline
\end{tabular}
\end{table*}

\subsection{Expected scatter in the correlation}

In the following we will study the correlations between the radio powers of the relic and the corresponding Mach numbers derived from X-ray and radio measurements. Given the large number of quantities that can impact on the radio luminosity of the relics (e.g., electrons density, magnetic field, size of the relic, properties of the seed electrons) we do not expect that the radio powers will follow strictly the trend as a function of $\mach$ that is predicted for the DSA and for the adiabatic compression models through the function $\Psi(\mach)$ and the amplification factor $A$, respectively. However, the dependence of $P_{1.4}$ is generally stronger from $\mach$ than from any other parameter involved, especially at low values of $\mach$. 

For example, Pinzke et al. (2013) calculated the scatter in the $P_{1.4}-\mach$ plane produced by the variation of the several parameters for the DSA model operating on the direct acceleration from the thermal gas and on a fossil cloud of relativistic electrons (see their fig.16), finding that the two models should predict a different slope of the average correlation for $\mach\le4$, and a very different number of relics for $\mach\le3$ even when reasonable variations of the other parameters are taken into account (see also Vazza \& Br\"uggen 2014). 

In order to extend this result also to the adiabatic compression scenario, we performed a Monte Carlo simulation of $10^5$ relics assuming an initial distribution of radio powers of the clouds following the one of the local radio luminosity function of radio loud AGNs (Mauch \& Sadler 2007), a distribution of the Mach numbers produced in internal shocks according to Ryu et al. (2003), a uniform distribution of the initial spectral indices in the range $s=2-5$, and applying the corresponding amplification factor as in eq.(\ref{amplif.compr}). We calculated the average correlation line as the best fit value with a power law to the simulated data, and the maximum curve as the one with the same slope including 95\% of the simulated data.

\begin{figure}
\centering
\begin{tabular}{c}
\includegraphics[width=85mm,height=65mm]{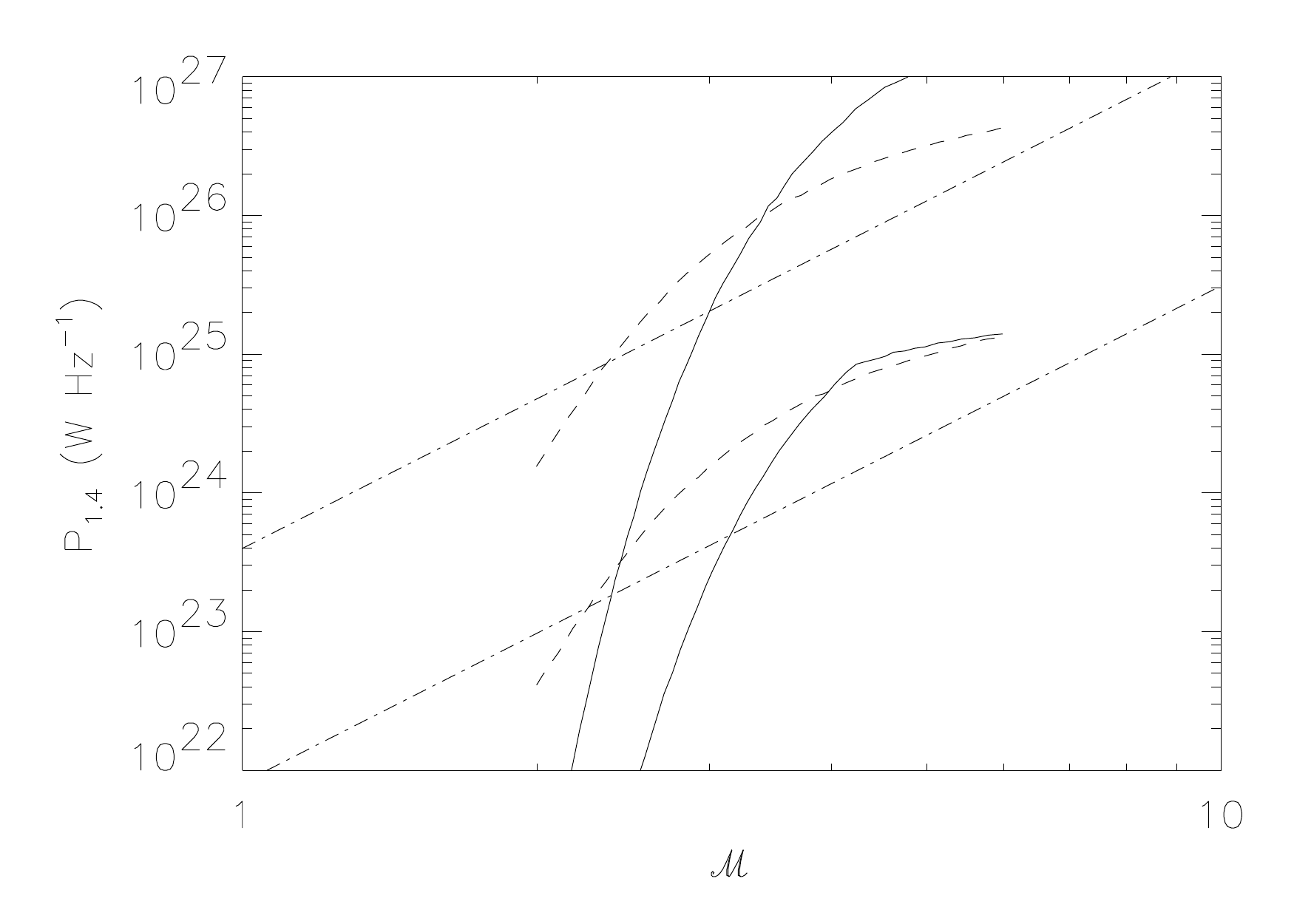}
\end{tabular}
\caption{Average and maximum correlation lines predicted by different models: DSA from the thermal pool (solid lines), DSA on a relativistic fossil cloud (dashed lines), and adiabatic compression on a relativistic fossil cloud (dot-dashed lines). The first two sets of curves are taken from Pinzke et al. (2013), the last one from our Monte Carlo simulation (see text for detail).}
\label{fig:models}
\end{figure}

Our result is compared with the curves taken from Pinzke et al. (2013) in Fig.\ref{fig:models} (these curves are shown in their plot only in the range $\mach=2-6$). We find that there is a difference between the three models especially at low Mach numbers: we tried to fit the curves with a power law in the range $\mach\le4$, and we found values of the slopes of the order of 10--11 for the direct DSA, of 7--9 for the DSA on a relativistic cloud, and $\sim3.6$ for the adiabatic compression.

In order to check the effect of using a small number of relics (in our case 20 relics) on the determination of the slope that can be obtained from a fit to the data, we also performed Monte Carlo simulations: for each of the three models shown in Fig.\ref{fig:models}, we simulated 20 relics allowing the several parameters to vary within reasonable values in galaxy clusters, according to eq.(\ref{eq:epr}) for the direct DSA, eq.29 in Pinzke et al. (2013) for the DSA on a relativistic cloud, and eqs.(\ref{eq.power.compr}-\ref{amplif.compr}) for the adiabatic compression, and fitted the resulting $P_{1.4}-\mach$ relation with a power-law $P_{1.4}=a\mach_X^b$. We repeated this calculation $10^4$ times for each model, and we found that the resulting slopes are following a distribution with $\bar b=8.8\pm2.0$ for the direct DSA, $\bar b=5.1\pm1.6$ for the DSA on a relativistic cloud, and $\bar b=3.6\pm1.0$ for the adiabatic compression.

Therefore in the following, for the largest sample of relics where the Mach number from X-ray observation has been estimated, we can try to identify from these correlations if the average trend is resembling more closely one of these scenarios. 
In particular, a change in the slope at Mach number $\mach \sim 3$ is typical of the DSA scenarios, and a continuous power-law increase is typical of the adiabatic compression scenario. Also, the presence of a high number of relics with high radio power and $\mach\simlt2$ could be explained only in terms of a scenario where the radio power is mainly determined from the initial cloud.

After the study of the $P_{1.4}-\mach$ correlation, we will also check the correlation between the values of the Mach number derived in different ways in order to determine which method is more reliable for the study of the properties of radio relics.

\subsection{Radio power vs. X-ray derived Mach numbers}

\begin{figure}
\centering
\begin{tabular}{c}
\includegraphics[width=85mm,height=65mm]{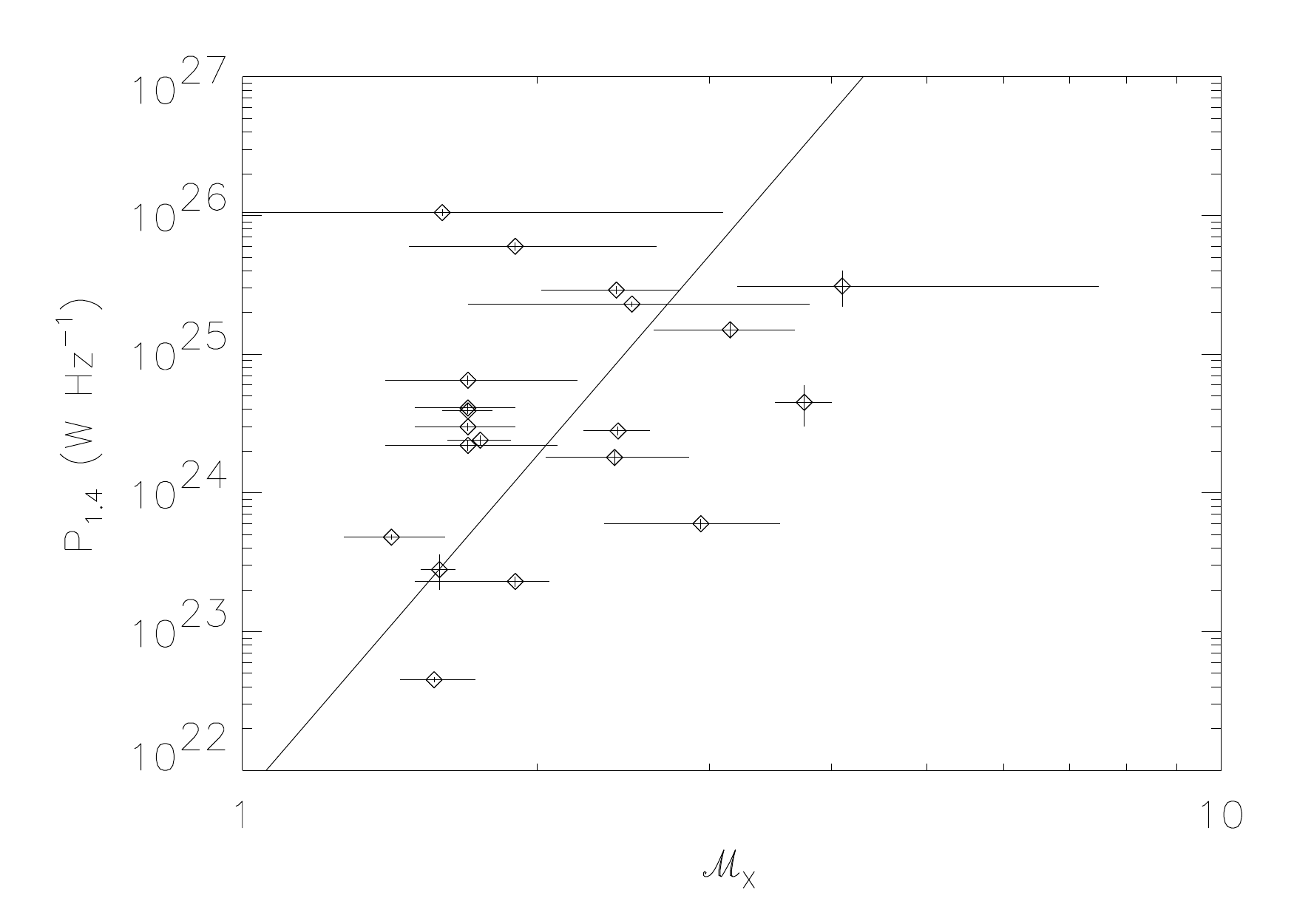}
\end{tabular}
\caption{Correlation between the radio power of the relic $P_{1.4}$ and the X-ray derived Mach number $\mach_X$. The best fit line corresponding to a power-law $P_{1.4}=a\mach_X^b$ is plotted for $\mbox{Log } a = 21.8\pm0.4$ and $b=8.2\pm0.9$.}
\label{fig:power_machx}
\end{figure}

We show in Fig.\ref{fig:power_machx}  the correlation between the power of the radio relic as measured at 1.4 GHz, $P_{1.4}$, and the X-ray derived Mach number $\mach_X$. For a power law relation, $P_{1.4}=a\mach_X^b$, the best fit is obtained for the values $\mbox{Log } a = 21.8\pm0.4$\footnote{Here and in the following we use the notation $\mbox{Log }x\equiv \mbox{log}_{10}(x)$.} and $b=8.2\pm0.9$, giving a minimum $\chi^2$ value of $\chi^2_{min}=100$ for 18 d.o.f. ($\chi^2_{red}=5.55$). Given the high value of $\chi^2_{min}$ this fit is therefore not good, as is clear given the large scatter of the data. 
We notice that the value of the slope indicated by the formal best fit is closer to the DSA case 
than to the adiabatic compression case, where the slope should be of the order of the spectral index of the electrons (see Figs.\ref{fig:amp} and \ref{fig:models}). For $s=8.2$, the radio spectral index should be $\alpha\sim(8.2-1)/2=3.6$, that is not a representative value for radio relics (see Table \ref{tab:Mach_X_radio}).
By comparing the value of the slope as resulting from this fit to the distribution of the results of the fits to a sample of 20 simulated relics as described in the previous subsection, and calculating the quantity
\begin{equation}
t=\frac{|b-\bar b|}{\sqrt{\sigma_b^2+\sigma_{\bar b}^2}} ,
\label{t.student}
\end{equation}
we can estimate that the probability that a $t$-distributed variable with 18 d.o.f. has a smaller value than the calculated one is 60.6\%, 94.6\%, and 99.8\% for the direct DSA, the DSA on a relativistic cloud and the adiabatic compression respectively.\\
In order to determine if a correlation between the data is at all present we calculated the Pearson $\rho$ coefficient for the correlation between $\mbox{Log }P_{1.4}$ and $\mbox{Log }\mach_X$, given by the formula
\begin{equation}
\rho=\frac{\sum_i(x_i-\bar{x})(y_i-\bar{y})}{\sqrt{\sum_i(x_i-\bar{x})^2}\sqrt{\sum_i(y_i-\bar{y})^2}}
\label{pearson}
\end{equation}
and we found that the value obtained for these sets of the data is $\rho=0.335$. 
The threshold value for 18 d.o.f. of $\rho$ in a two-tails test over which the difference from an absence of correlation is significant at 95\% confidence level is 0.444; therefore we can conclude that no significant correlation is present between these two sets of variables.
In order to take into account the possible effect of the error bars on the value of the Pearson coefficient, we performed Monte Carlo simulations, by allowing each data point to vary around its central value with a Gaussian distribution with the observed value of $\sigma$. We simulated $10^6$ distributions, and we found that the probability that the Pearson coefficient is in the region of no correlation is 79.9\%.
This indicates that the data do not show any
evidence of correlation when both their central values and their error bars are considered. We will discuss in Sect.4 the consequences of this result.

\subsection{Radio power vs. radio derived Mach numbers}

\begin{figure}
\centering
\begin{tabular}{c}
\includegraphics[width=85mm,height=65mm]{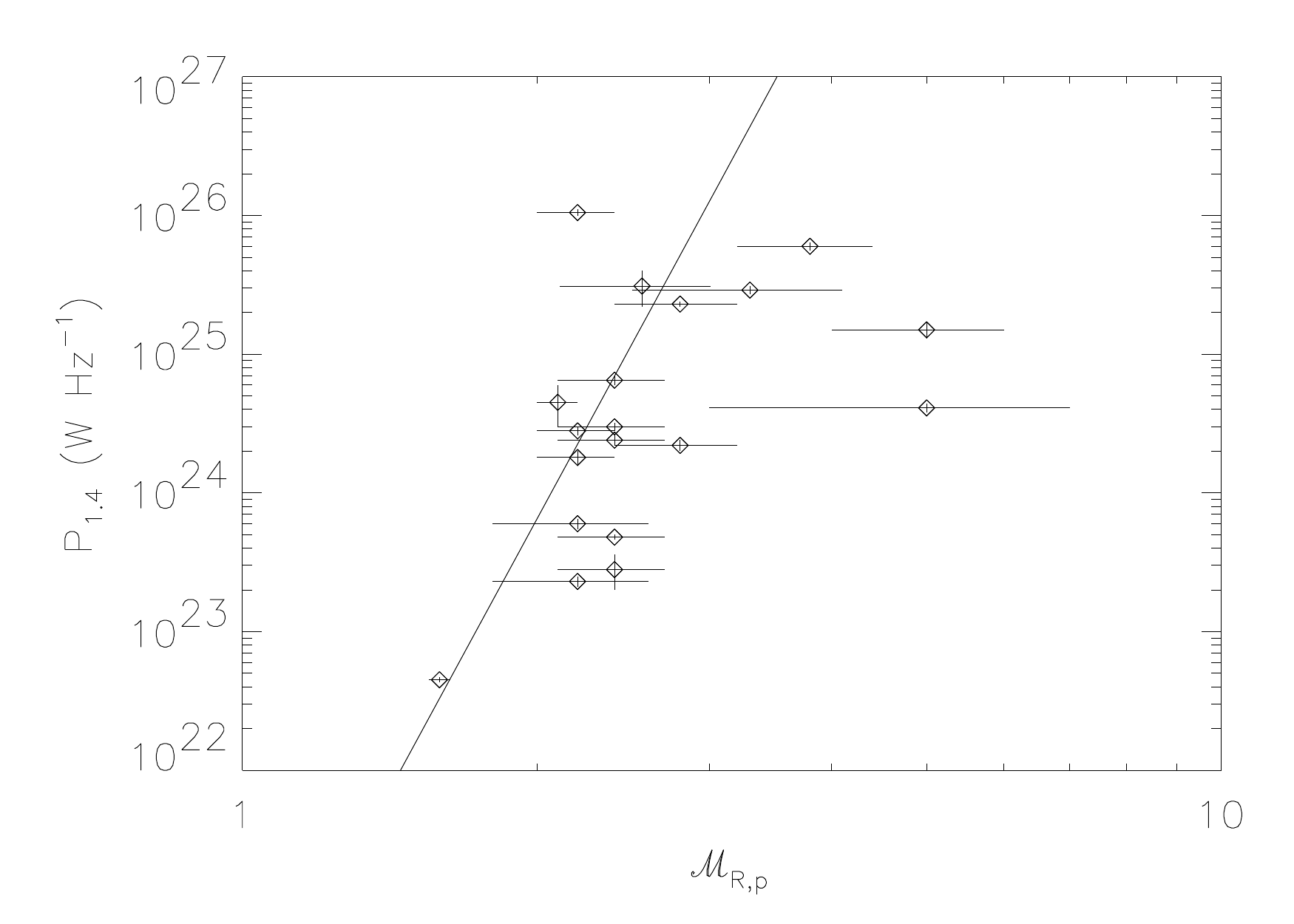}
\end{tabular}
\caption{Correlation between the radio power of the relic $P_{1.4}$ and the Mach number derived from the peak value of the radio spectral index $\mach_{R,p}$. The best fit line corresponding to a power-law $P_{1.4}=a\mach_{R,p}^b$ is plotted for $\mbox{Log } a = 19.9\pm0.4$ and $b=13\pm1$.}
\label{fig:power_machrp}
\end{figure}

We show in Fig.\ref{fig:power_machrp}  the correlation between the power of the radio relic $P_{1.4}$ and the Mach number derived from the peak value of the radio spectral index $\mach_{R,p}$. For a power law relation, $P_{1.4}=a\mach_{R,p}^b$, the best fit is obtained for the values $\mbox{Log } a = 19.9\pm0.4$ and $b=13\pm1$, giving a minimum value $\chi^2_{min}=45.8$ for 17 d.o.f. ($\chi^2_{red}=2.69$). 
Again the $\chi^2_{min}$ value is quite high indicating that this fit is not good and seems to be biased towards reproducing low $\mach$ numbers, as is evident from the large scatter of the data and the large errors for high $\mach$ values.
The comparison with the $\bar b$ values as found in Sect.3.2 shows that the probability that a $t$-distributed variable has a smaller value than the one derived from eq.(\ref{t.student}) using this best fit value is 96.2\% for the DSA case, and bigger than 99.9\% for the DSA on relativistic clouds and the adiabatic compression cases.
The Pearson correlation coefficient calculated according to eq.(\ref{pearson}) is $\rho=0.510$; for 17 d.o.f. the threshold value at 95\% c.l. is 0.455; therefore in this case a correlation is present between the two sets of data. However, the results of the Monte Carlo simulation taking into account the effect of the error bars give the result that the probability of no correlation is 54.2\%; therefore, in this case the error bars don't allow to draw a definite conclusion.

\begin{figure}
\centering
\begin{tabular}{c}
\includegraphics[width=85mm,height=65mm]{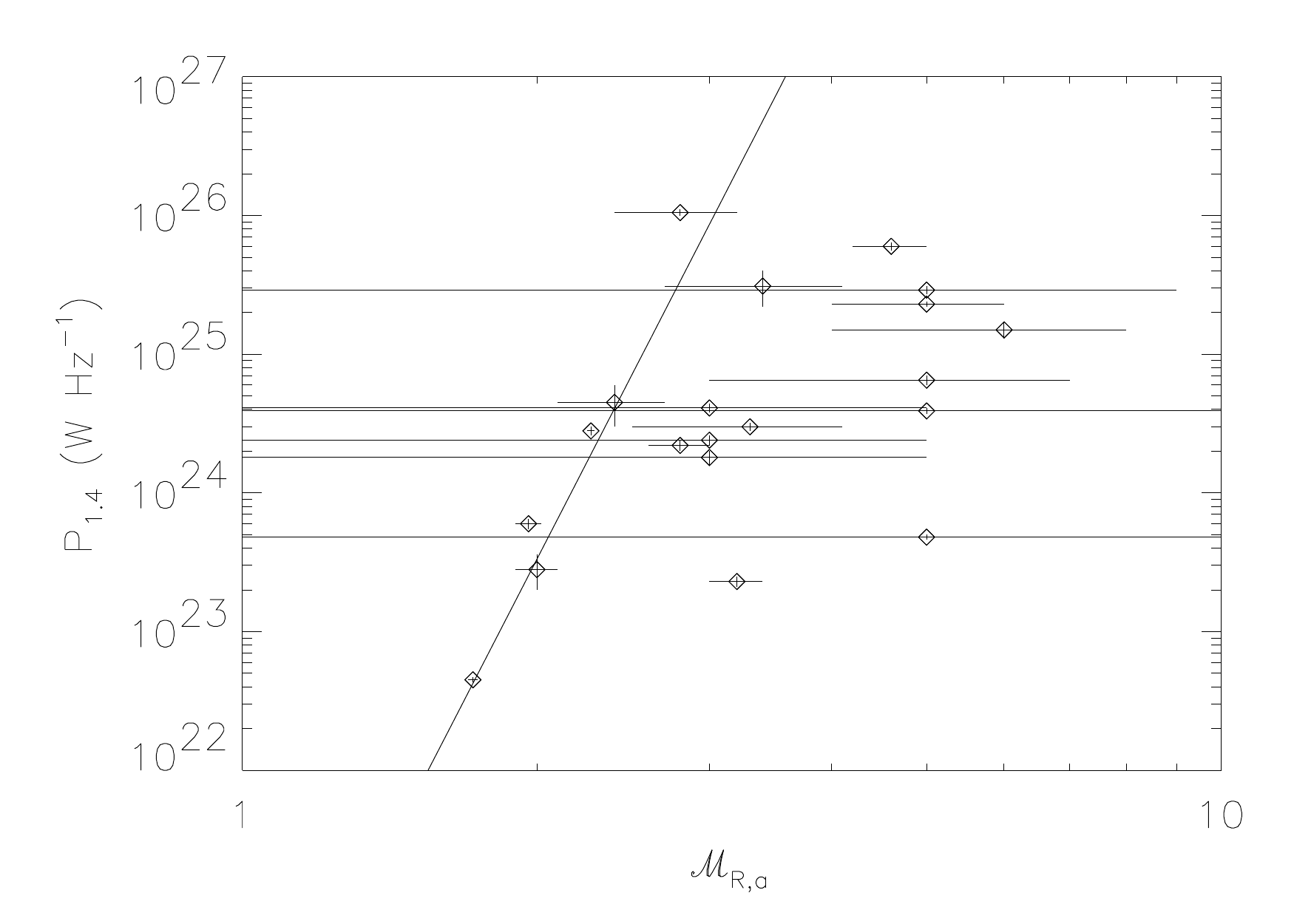}
\end{tabular}
\caption{Correlation between the radio power of the relic $P_{1.4}$ and the Mach number derived from the average value of the radio spectral index $\mach_{R,a}$. The best fit line corresponding to a power-law $P_{1.4}=a\mach_{R,a}^b$ is plotted for $\mbox{Log } a = 19.4\pm0.1$ and $b=13.7\pm0.7$.}
\label{fig:power_machra}
\end{figure}

We show in Fig.\ref{fig:power_machra} the correlation between the power of the radio relic $P_{1.4}$ and the Mach number derived in this case from the average value of the radio spectral index $\mach_{R,a}$. For a power law relation, $P_{1.4}=a\mach_{R,a}^b$, the best fit is obtained for the values $\mbox{Log } a = 19.4\pm0.1$ and $b=13.7\pm0.7$, giving a minimum value $\chi^2_{min}=132$ for 18 d.o.f. ($\chi^2_{red}=7.33$). 
As in the previous case, the $\chi^2_{min}$ value is quite high indicating that this fit is not good and seems to be biased towards reproducing low $\mach$ numbers, as is evident from the large scatter of the data and the large errors for high $\mach$ values.
The comparison with the $\bar b$ values as found in Sect.3.2 shows that the probability that a $t$-distributed variable has a smaller value than the one derived from eq.(\ref{t.student}) using this best fit value is 98.4\% for the DSA case, and bigger than 99.9\% for the DSA on relativistic clouds and the adiabatic compression cases.
The Pearson correlation coefficient calculated according to eq.(\ref{pearson}) is $\rho=0.551$; for 18 d.o.f. the threshold value at 95\% c.l. is 0.444, indicating therefore also in this case a correlation is present between the two sets of data. However, the effect of the error bars is to indicate that the probability of no correlation is 81.2\%; the discrepancy between these results is due to the big size of some of the error bars, as evident from Fig.\ref{fig:power_machra}.

\subsection{X-ray vs. radio derived Mach numbers correlation}

\begin{figure}
\centering
\begin{tabular}{c}
\includegraphics[width=110mm,height=70mm]{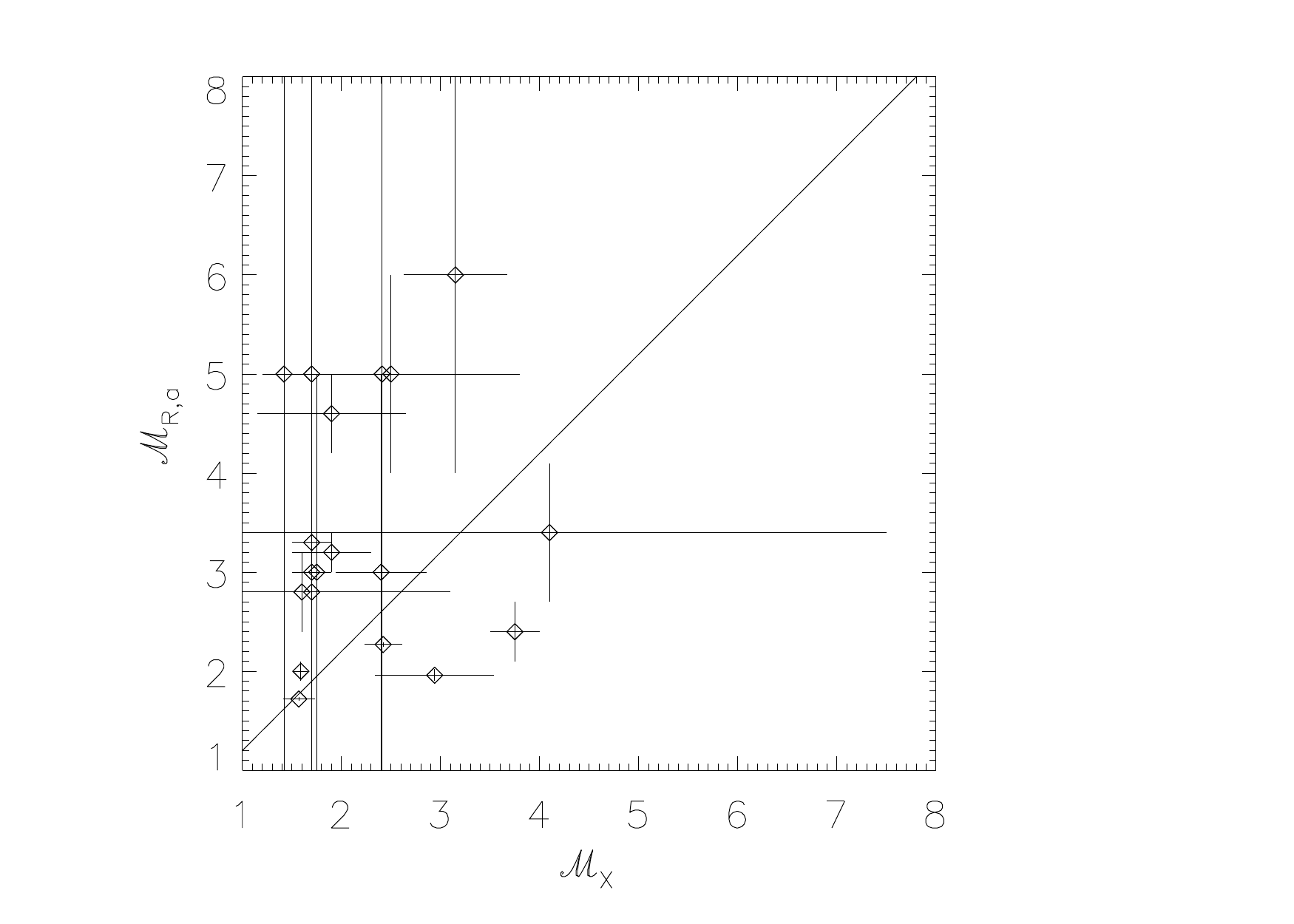}
\end{tabular}
\caption{Correlation between the Mach number derived from the average value of the radio spectral index $\mach_{R,a}$ and from X-ray measurements. The best fitting line $M_{R,a}=a+b\mach_X$ is plotted for $a=0.2\pm0.4$ and $b=1.0\pm0.2$.}
\label{fig:machx_machra}
\end{figure}

\begin{figure}
\centering
\begin{tabular}{c}
\includegraphics[width=110mm,height=70mm]{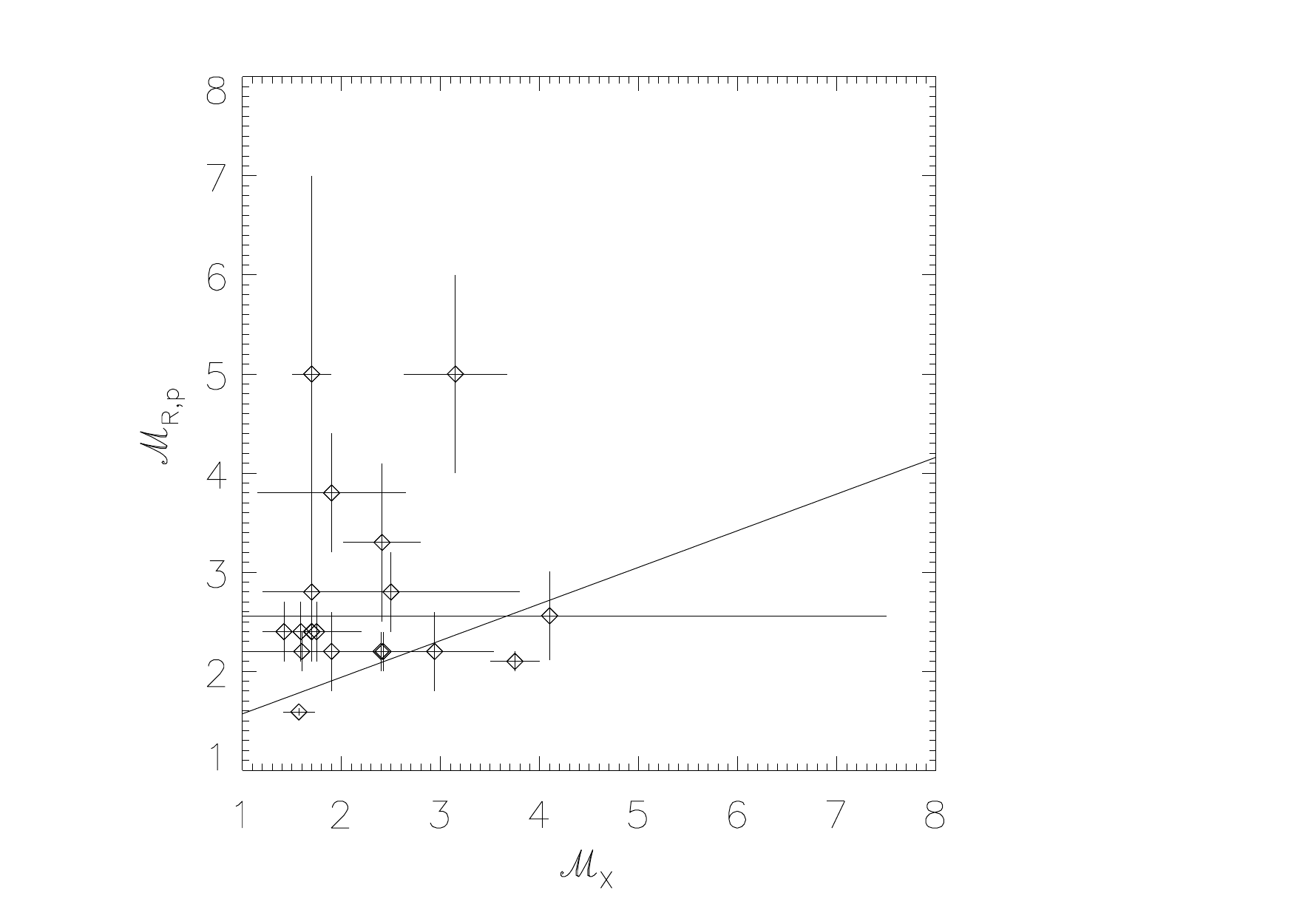}
\end{tabular}
\caption{Correlation between the Mach number derived from the peak value of the radio spectral index $\mach_{R,p}$ and from X-ray measurements. The best fitting line $M_{R,p}=a+b\mach_X$ is plotted for $a=1.2\pm0.2$ and $b=0.37\pm0.08$.}
\label{fig:machx_machrp}
\end{figure}

\begin{figure}
\centering
\begin{tabular}{c}
\includegraphics[width=110mm,height=70mm]{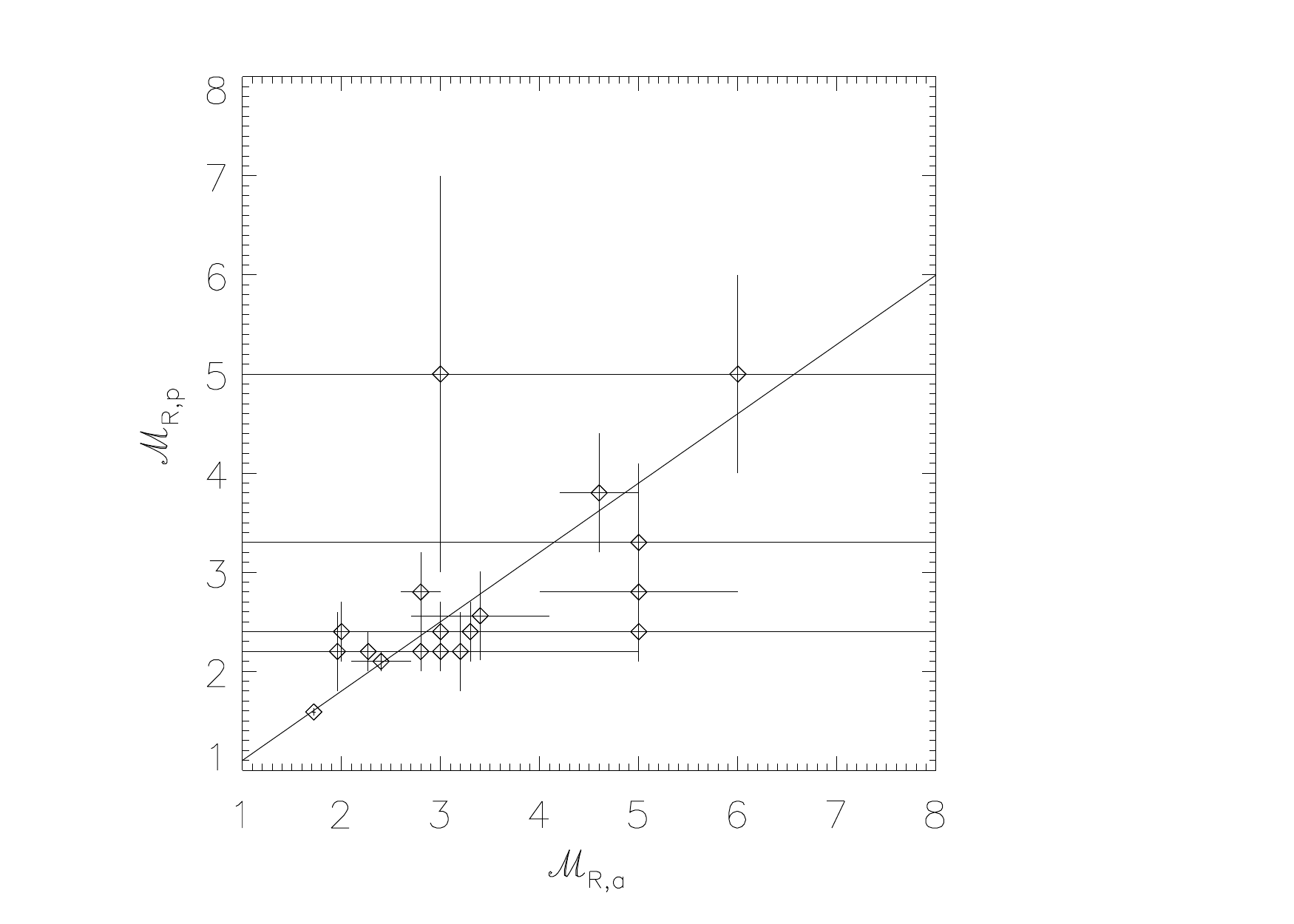}
\end{tabular}
\caption{Correlation between the Mach number derived from the average value of the radio spectral index $\mach_{R,a}$ and from the peak value of the radio spectral index $\mach_{R,p}$. The best fitting line $M_{R,p}=a+b\mach_{R,a}$ is plotted for $a=0.4\pm0.2$ and $b=0.7\pm0.1$.}
\label{fig:machrp_machra}
\end{figure}

We show in Figs. \ref{fig:machx_machra},  \ref{fig:machx_machrp} and \ref{fig:machrp_machra} the correlations between the Mach number derived from the average value of the radio spectral index $\mach_{R,a}$ and from X-ray measurements, the one between the Mach number derived from the peak value of the radio spectral index $\mach_{R,p}$ and from X-ray measurements as well as the correlation between the Mach number derived from the average value of the radio spectral index $\mach_{R,a}$ and from the peak value of the radio spectral index $\mach_{R,p}$, respectively.\\
We fitted (see Fig.\ref{fig:machx_machra}) the relation between $\mach_{R,a}$ and $\mach_X$ with a linear function $M_{R,a}=a+b\mach_X$, and found that the best fit is obtained for $a=0.2\pm0.4$ and $b=1.0\pm0.2$ with $\chi^2_{min}=54.8$ for 18 d.o.f. ($\chi^2_{red}=3.04$). Therefore, the parameters of the fit are compatible with the $\mach_{R,a}=\mach_X$ condition, but the fit is poor. The Pearson coefficient for these two sets of data is $\rho=0.011$, whereas the threshold value at 95\% confidence level is 0.444; therefore, a correlation is not present between the two sets of data, as confirmed also by the Monte Carlo simulations taking into account the error bars, that indicate a 99.8\% of probability of not having a correlation.

For the relation between $\mach_{R,p}$ and $\mach_X$ we found (see Fig.\ref{fig:machx_machrp}) that the best fitting parameters are $a=1.2\pm0.2$ and $b=0.37\pm0.08$ with $\chi^2_{min}=65.3$ for 17 d.o.f. ($\chi^2_{red}=3.84$). Therefore in this case the fitting is worse than in the previous case, and the best parameters are biased towards a value of $a$ different from 0 and a slope $b$ smaller than 1. The Pearson coefficient for these sets of data is 0.079, and the threshold value is 0.455, therefore no correlation is present between these sets of data, as confirmed also by the Monte Carlo simulations taking into account the error bars, that indicate a 99.6\% of probability of not having a correlation.

Finally, for the relation between $\mach_{R,p}$ and $\mach_{R,a}$ we found (see Fig.\ref{fig:machrp_machra}) that the best fitting parameters are $a=0.4\pm0.2$ and $b=0.7\pm0.1$ with $\chi^2_{min}=13.2$ for 17 d.o.f. ($\chi^2_{red}=0.776$). Therefore, in this case the fit is good, but the best fitting parameters are not compatible with the relation $\mach_{R,p}=\mach_{R,a}$ until $\sim3\sigma$ level. The Pearson coefficient is 0.539, and the threshold value is 0.455, therefore in this case a correlation is present between these two sets of data. However, the probability of no correlation obtained from the Monte Carlo simulations is 79.6\%; even in this case, the discrepancy between the results obtained by considering only the central values and by considering the error bars is due to the big size of some of the error bars.

These results show that the determination of the Mach number associated to a shock is a difficult task. While $\mach_X$ is definitely associated to the properties of the shock, the Mach numbers derived from radio measures are really determined from the spectral index of the relic under the assumption of DSA, and therefore do not necessarily reflect the properties of the shock if the relic is produced by a different mechanism, like electrons ejected from radiogalaxies. Generally speaking, the lack of correlation between $\mach_X$ and $\mach_R$ (in both the two ways this second quantity can be determined), is therefore a clue that the spectral index of the relic is not due to the acceleration of particles by the shock but to the fossil electrons properties. More specifically, from both the correlations between $\mach_{R,a}$ and $\mach_{R,p}$ with $\mach_X$, it seems that there is not any correlation for $\mach_X < 3$, where the radio determined Mach numbers are generally higher than the X-ray determined ones, according to the results published in literature (see, e.g., Akamatsu et al. 2017 for the most recent compilation and analysis). This means that for low Mach numbers the spectrum of the relics is usually flatter than what is expected from the DSA, indicating that we are probably observing the properties of a fossil electron population. Instead, for $\mach_X>3$, it seems that the radio determined Mach numbers are lower than the X-ray determined values. This means that the spectra of relics for strong Mach numbers are steeper than the values expected from the DSA, and this is very interesting, because in this case the acceleration process is expected to determine the slope of the spectrum of the relic even if the electrons are a fossil population that is being reaccelerated (e.g. Kang \& Ryu 2011). Therefore, this can be a clue that a different process is working to determine the relic spectrum, as the adiabatic compression scenario, that is expected to generate electrons spectra steeper than DSA (En\ss lin \& Gopal-Krishna 2001). However, the small number of points with $\mach_X>3$ and the large error bars associated to high Mach numbers do not allow to derive yet a definite conclusion. 

The only case where there is a hint of correlation is for $\mach_{R,a}$ vs. $\mach_{R,p}$, according to their central values; in this case such a correlation is expected because both these quantities are determined from the spectral index of the relics. However, Fig.\ref{fig:machrp_machra} shows that in this case there is a good correlation for low Mach numbers up to values $\mach_{R,a}\simgt3$, but for higher Mach numbers the points are more scattered, and the error bars are so big to obscure the correlation. In particular, there is a number of points with $\mach_{R,a}>\mach_{R,p}$, i.e. where the average spectral index is flatter than the value that can be expected from its peak value. In the DSA scenario we should expect $\mach_{R,a}=\mach_{R,p}$, because the average and peak values of the spectral index are related by $\alpha_{av}=\alpha_{peak}+0.5$ because of the energy losses, while in the adiabatic compression scenario we should expect $\mach_{R,a}<\mach_{R,p}$, because the electrons should loose quickly their energy, because no acceleration processes should be effective and the electrons are outside the equilibrium condition. Therefore, the $\mach_{R,a}>\mach_{R,p}$ condition should imply the existence of other mechanisms, as for example the turbulences scenario (Fujita et al. 2015), that can accelerate the electrons in a wider area than the shock front one, allowing to flatten the spectral index in a large fraction of the relic area, and producing in this way an average value of the spectral index which is flatter than the one expected from the energy losses of the electrons. The fact that this trend is observed only for $\mach>3$ can instead suggest that this effect is due to the fact that when the shock wave is slow we can observe the electrons far from the shock front when they already have reached the equilibrium state, whereas when the shock wave is faster we can observe a larger area far from the shock front where the electrons did not have the time to reach the equilibrium state, and therefore their spectral index is flatter than the equilibrium one. 
However, also in this case the uncertainties in the determination of the radio Mach number for high values, and the small number of points, do not allow to derive a definite conclusion yet.

\section{Discussion}

The results of our analysis of the $P_{1.4}-\mach$ correlation set interesting constraints on the available models for the origin of radio relics. 
Our results are challenging for the available models and leave several open questions: in the $P_{1.4}-\mach_X$ distribution (Fig.\ref{fig:power_machx}) the observed slope is compatible with the direct DSA scenario, is marginally compatible ($\simlt2\sigma$) with the DSA on relativistic clouds scenario, and is not compatible ($\simgt3\sigma$) with the adiabatic compression scenario.
However, the data do not show a change in the slope of the distribution as expected in the DSA scenarios (see Fig.\ref{fig:models}). Moreover, 
this is in tension with
the presence of a number of relics with $\mach_X\simlt2$ and $P_{1.4} > 10^{24}$ W Hz$^{-1}$, that is expected only in the adiabatic compression scenario (see Fig.\ref{fig:models}), but with only a moderate contribution from the compression, because for $\mach_X\simlt2$ the compression factor is expected to be of one order of magnitude or less (see Fig.\ref{fig:amp}). This should imply that the luminosity of the relic is mainly determined from the properties of a fossil electron population rather than the properties of the shock. In the $P_{1.4}-\mach_R$ distributions (Figs.\ref{fig:power_machrp}--\ref{fig:power_machra}) the observed slope is marginally compatible with the direct DSA scenario (at $\simgt2\sigma$), and not compatible ($>3\sigma$) with the other two scenarios. More, the distribution 
$P_{1.4}-\mach_{R,p}$ (Fig.\ref{fig:power_machrp}) shows a large value of the slope until $\mach_{R,p}\sim3$, and the presence of a possible flattening after this value, suggested by the presence of some data points (actually, only three points) that are located significantly under the best-fit line, as expected in the DSA scenarios.  However, it is important to note that the values of $\mach_{R,p}$ are determined under the DSA assumption, and the fact that a similar trend is not observed using the $\mach_X$ values, which are not based on the DSA assumption, make this result weaker.

Also the correlation analysis provides results that are not simply explainable with the current models. In fact, there is a clear lack of correlation between $P_{1.4}$ and $\mach_X$ by considering both the central values of the data and the corresponding error bars. This result puts in difficulty all the considered models where the radio power should strongly depend on the Mach number, even when the effect of the other parameters affecting the radio power is considered (see Fig.\ref{fig:models}). Instead, data show that
there is a possible positive correlation between $P_{1.4}$ and the radio determined Mach numbers, when only the central values of the data are taken into account.  
However, this trend is not confirmed by the analysis of the correlation when the error bars are taken into account.
For this reason, better determinations of the radio spectral index are desirable in order to reduce the errors on $\mach_R$ and have results that can be compared with the ones derived from the X-ray derived Mach numbers.

If the properties of the relics are mainly related to a fossil electron population, the properties of the shock should become more evident at high frequencies (e.g. Macario et al. 2011) and at high Mach numbers (e.g. Kang \& Ryu 2011). Currently the observations of radio relics at high frequencies are contradictory, because from some observations it seems that a quick steepening of the spectrum is present around 10 GHz (Stroe et al. 2016), therefore favoring the adiabatic compression scenario, whereas from other observations it seems that no steepening is present until 30 GHz (Kierdorf et al. 2017), therefore favoring an acceleration (or re-acceleration) scenario. Therefore, the study of radio relics with high-sensitivity instruments at high frequencies like the SKA1-MID will be important to determine which astrophysical process is producing the relics. The determination of high values of Mach numbers from radio observations is difficult because the Mach number function tends to diverge for $\alpha_{peak}\rightarrow0.5$ and $\alpha_{av}\rightarrow 1$, so that even a small error in the determination of the spectral index can produce a huge error on the derived Mach number. Also in this case instruments with high sensitivity and high angular resolution (that can identify the regions where the spectral index has a peak) will be useful to better constrain the Mach number values derived from radio observations.

The correlation analysis between the Mach numbers derived using different techniques provides also contradictory results: while for $\mach_X<3$ it seems that no correlation with $\mach_R$ is present, for $\mach_X>3$ it seems that the values of $\mach_R$ are lower than expected, corresponding to steeper spectral index, therefore disfavoring scenarios based on acceleration or re-acceleration. Instead, from the correlation between $\mach_{R,p}$ and $\mach_{R,a}$ it seems that for $\mach_R>3$ the values of $\mach_{R,a}$ are higher than what is expected from $\mach_{R,p}$, suggesting that other forms of acceleration, produced by e.g. turbulences, are effective. The possible effect of turbulent reacceleration is in fact to produce spectral indices flatter than the ones predicted by the standard DSA, and this effect should be visible in a region behind the shock, displaced from the shock front by a distance of the order of 100--200 kpc (e.g. Fujita et al. 2015), therefore impacting on $\mach_{R,a}$ but not on $\mach_{R,p}$.

A possible alternative is that the shock is not produced in the ICM like in the case of merger or accretion shocks, but from other radiogalaxies jets (e.g., Kraft et al. 2012): in this case all the formulae used to determine the Mach number must be changed, and the value of the polytropic index $\gamma_g=4/3$, corresponding to a full relativistic plasma, should be used instead of the value of 5/3 typical of a thermal plasma. In this way, it would be also possible to obtain a minimum value of the peak spectral index $\alpha_{peak}=0.25$ (and also the formulae to obtain $\mach_R$ and $\mach_X$ will change); this would allow to explain, e.g., the peak values $\alpha_{peak}\sim0.4$ for the relics in A3667 observed by Hindson et al. (2014), and the average value of $\alpha_{av}=0.8\pm0.1$ found in A2255 by Pizzo \& De Bruyn (2009). Therefore, it is necessary to study in detail the spatial properties of the (radio)galaxies close to the relics, and check if both the electrons and the shocks can be produced by radio galaxies jets (we will discuss this case more extensively elsewhere, see Colafrancesco et al. in preparation).

Other possibilities that have been proposed in literature to explain the lack of correspondence between the properties of the relics and the Mach numbers derived with different techniques include: shocks with non uniform distribution of the Mach numbers due to ICM inhomogeneities (e.g., Nagai \& Lau 2011), the effect of the dynamical reaction of particles on the shock (e.g. Blasi 2010), pre-acceleration of electrons via shock-drift acceleration (Guo, Sironi \& Narayan 2014), projection effects (Skillman et al. 2013), non-equilibrium effects between electron and proton temperatures after the shock (Akamatsu et al. 2017).

\section{Conclusions}

We have discussed in this paper a new technique to constrain models for the origin of radio relics in galaxy cluster which makes use of the correlation between the shock Mach number and the radio power of relics. This analysis has been carried out using a sample of relics with information on the Mach numbers derived using various techniques: the Mach number derived from X-ray observation of shock jump conditions and those derived using spectral information from radio observations using the the peak and the average values of the spectral index along the relic. 

The lack of correlation between $\mach_X$ and $\mach_R$  is an indication that the spectral index of the relic is likely not due to the acceleration of particles operated by the shock but it is related to the properties of a fossil electrons population (see Sect.3.5). The data show that at $\mach_X\simlt3$ there is a trend  $\mach_R>\mach_X$, as found in previous studies (e.g. Akamatsu et al. 2017), showing that the strength of the shock is not able to produce the observed flatness of the spectrum, whereas for $\mach_X>3$ there is a trend pointing towards $\mach_R<\mach_X$, again suggesting that a process different from DSA is producing the observed spectral index. However, the small number of data points with $\mach_X>3$ and the large error bars on $\mach_R$ do not allow to derive definite conclusions yet. Therefore, an increase of the number of observed relics with precise measurements of $\mach_R$ and with relatively small uncertainties, especially for high values of $\mach_R$, as feasible with new generation instruments like SKA1-MID, will be important to determine the consistency between the physics of shock as described by jump conditions in gas density and temperature (as measured in the X-rays) and the effects of shocks in the possible (re)energization of electrons  by DSA or adiabatic compression models. The trend $\mach_{R,a}>\mach_{R,p}$ observed for $\mach_{R,a}>4$ is instead suggesting that some other mechanism, like, e.g., the turbulences scenario, are working.

The available data on the correlation between the radio power $P_{1.4}$ and Mach numbers ($\mach_R$ and $\mach_X$) in relics indicate that neither the DSA nor the adiabatic compression can simply reproduce the observed $P_{1.4}-\mach$ correlations.
In fact, while the slope of the $P_{1.4}-\mach_X$ correlation and a possible flattening of the $P_{1.4}-\mach_{R,p}$ correlation at high $\mach_{R,p}$ values seem to point towards a DSA scenario, the lack of a similar flattening in the $P_{1.4}-\mach_X$ plot and the relatively high number of relics with low Mach number and high radio power are instead typical of the adiabatic compression scenario, suggesting that the relic radio power is mainly determined by the properties of a fossil electron population. 
This is confirmed also by the lack of correlation between $P_{1.4}$ and $\mach_X$, whereas there is a possible positive correlation between $P_{1.4}$ and the radio determined Mach numbers, when only the central values of the data are taken into account. However, this trend is not confirmed by the analysis of the correlation when the error bars are taken into account.

The results we obtained in this paper require either to consider models of shock (re)acceleration that go beyond the proposed scenarios of DSA and adiabatic compression at shocks, or lead to reconsider the origin of radio relics: along this last avenue, we suggest that a necessary step forward is provided by the detailed study of the evolution of seed electron populations as those produced by radio galaxy jets and outflows. This analysis must be done for each specific relic using multi-messenger techniques and we will describe the results of our analysis of the 77 relics found in our extended cluster relic sample in a forthcoming paper (Colafrancesco et al., in preparation).

Additional information in the radio band, especially regarding the spectral index at high frequencies which can be obtained with instruments like SKA1-MID, will be useful to better constrain the properties of radio relics and the models for their origin. Information obtained in other spectral bands will be useful to complete this analysis. For example, using X-ray instruments with high sensitivity and spatial and spectral resolution, like Athena or Astro-H2, it would be possible to determine $\mach_X$ with better precision.
From measurements in the mm. band the Mach number can be obtained in an independent way from the pressure jump estimated from the thermal Sunyaev-Zel'dovich measurements (e.g. Erler et al. 2015), and possible detection of the non-thermal SZE can give information on the properties of the electrons which are not depending on the magnetic field properties (see e.g. Colafrancesco \& Marchegiani 2011, Colafrancesco et al. 2013 for the case of radio galaxies lobes). In the hard X-ray and gamma rays, the possible detection of the Inverse Compton emission from relativistic electrons in relics can constrain their high-energy spectra, allowing to discriminate between the acceleration and adiabatic compression scenario (see Paulo, Colafrancesco \& Marchegiani 2017).\\
Therefore, the results presented in this paper show that deep and systematic spectral and imaging radio, mm. and high-energy observations of radio relics are essential to unveil the origin of this cosmic phenomenon which is a prototype for the evolution of non-thermal phenomena in galaxy clusters.

%%%%%%%%%%%%%%%%%%%%%%%%%%%%%%%%%%%%%%%%%%%%%%%%%%%%

\section*{Acknowledgments}
We thank the Referee for several useful comments and suggestions.
S.C. acknowledges support by the South African Research Chairs Initiative of the Department of Science and Technology and National Research Foundation and by the Square Kilometers Array (SKA). P.M. and C.M.P.  acknowledge support from the DST/NRF SKA post-graduate bursary initiative.
We thank A. Kudoda for his help in performing Monte Carlo simulations.

%%%%%%%%%%%%%%%%%%%%%%%%%%%%%%%%%%%%%%%%%%%%%%%%%

%%%%%%%%%%%%%%%%%%%%%%%%%%%%%%%%%%%%%%%%%%%%%%%%%%%%%%%%

\bsp

\label{lastpage}

\end{document}